\documentstyle[12pt]{article}
\begin{document}
\bibliographystyle{plain}
\newcommand{\be}{\begin{equation}}
\newcommand{\ee}{\end{equation}}
\newcommand{\bea}{\begin{eqnarray}}
\newcommand{\eea}{\end{eqnarray}}
\newcommand{\ddl}[1]{\stackrel{\leftarrow}{\partial\over\partial#1}}
\newcommand{\ddr}[1]{\stackrel{\rightarrow}{\partial\over\partial #1}}
\newcommand{\ddd}[2]{{\partial^2\over\partial #1\partial #2}}
\title
{Weyl Expansion for Symmetric Potentials}

\author{B. Lauritzen and N. D. Whelan\\
Niels Bohr Institute\\
Blegdamsvej 17, DK-2100, Copenhagen \O, Denmark}
\date{\today}
\maketitle

\begin{abstract}
We present a semiclassical expansion of the smooth part of the density
of states in potentials with some form of symmetry.
The density of states of each irreducible representation is separately
evaluated using the Wigner transforms of the projection operators.
For discrete symmetries the expansion yields a formally exact but
asymptotic series in $\hbar$, while for the rotational $SO(n)$ symmetries
the expansion requires averaging over angular momentum as well as energy.
A numerical example is given in two dimensions, in which we calculate
the leading terms of the Weyl expansion as well as the leading periodic
orbit contributions to the symmetry reduced level density.
\end{abstract}

\newpage


\section{Introduction}

In the analysis of quantum spectra it is common to decompose the full density
of states into a smooth average term with a superimposed fluctuation,
$\rho(E)=\bar{\rho}(E) + \rho_{\rm osc}(E)$. Each of
these contributions has a distinct interpretation in terms of classical
mechanics. The average or Weyl term is, to first order, understood as arising
from orbits of zero
length and is proportional to the allowed volume of classical phase
space, while the oscillating term reflects the periodic orbit structure
of the system.

The smooth part of the density of states can be obtained semiclassically,
by expanding the density operator $\hat{\rho} = \delta (E-\hat{H})$
as a series in $\hbar$,
\be \label{weylll}
\rho (E)={\rm tr}\hat\rho = \sum_n A_n(E) \hbar^n
\ee
which we shall refer to as the Weyl expansion.
Formally this gives an exact expansion of $\rho$,
and we should recover from this series both the average and the
oscillating parts of the level density.
On the other hand, any truncation of Eq.~(\ref{weylll}) is only a slowly
varying function of the energy and does not contribute an oscillating term to
the level density.
The resolution is that in general we expect the Weyl expansion to be
asymptotic in $\hbar$ and it has been
argued \cite{Berry} that information about the oscillating part of the
density is encoded in the divergent part of the series.
Indeed, by resumming the remainder of the series in a billiard system
oscillatory contributions given by the shortest  periodic orbits are
found.

%

The purpose of this paper is to expand the Weyl term in the presence of
either a discrete or a continuous symmetry. Such a symmetry implies
conserved quantum numbers so that the density decomposes
into a sum of partial densities, $\rho (E)=\sum \rho_\alpha(E)$,
where $\alpha$ labels a complete or partial list of conserved quantum
numbers.
We are more often than not interested in the reduced densities
$\rho_\alpha(E)$, rather than the full density of states.
For example, a quantum system composed of identical particles
(e.g. fermions) have wavefunctions that are totally odd or even under
particle exchange and the density of states is given by the density of
the antisymmetric or symmetric representations of the symmetric
group.
Another example would be an analysis of an
experiment in the presence of selection rules. For instance, we might have
an experimental probe of a spherically symmetric potential which is only
sensitive to states with a specific value of $J$, so we would need
to know the density of such states.

A practical use of the Weyl expansion is in
measuring the reliability of both experiments and numerical calculations.
By comparing the
reported density of states to the Weyl expansion one can decide if one
has, for instance, missed levels. In the presence of symmetry, it is
easier to do experiments and numerical calculations one symmetry representation
at a time and we need the symmetry reduced Weyl
expansion for comparison.  Finally, it is common to look at various
statistical aspects of spectra such as spacing distributions \cite{Bohigas}
and to do so
requires that we first unfold the spectrum by removing the Weyl term from the
density of states.  Since these statistics are only meaningful when computed
separately for each symmetry representation, it is necessary to know the
symmetry decomposed Weyl expansion.

As an asymptotic series, the Weyl expansion is most accurate when
truncated one term before
its least term and therefore the number of terms which we are allowed to
include in the smooth part depends on $\hbar$.
We will not go into any details on resumming the Weyl expansion.
Rather we adopt the practical point of view that we can simply add the
leading oscillatory part of the density to the smooth density of states.
Thus, for the partial density $\rho_\alpha$ of states belonging to the
irreducible representation $\alpha$ of the symmetry group we write
\be
\rho_\alpha (E) \approx {\bar \rho}_\alpha(E) +\rho_{\alpha,{\rm osc}}(E)
\ee
where the average term is given by the truncated Weyl expansion.
The oscillating term $\rho_{\alpha,{\rm osc}}$ has previously been
treated for both discrete [3-5]
and continuous \cite{sccthes,CL} symmetries.
To first order it yields a trace formula similar to the Gutzwiller
trace formula but with the periodic orbits given in a reduced phase-space.

The Weyl expansion in the presence of discrete symmetries was previously
discussed in
Refs.~[8-10].
Our formalism is the same as that
of Refs.~\cite{BTU,weidenmuller2}. In Ref.~\cite{BTU} however, the discussion
was limited to a single or several commuting reflections while
Ref.~\cite{weidenmuller2} was specific to the symmetric group
for identical particles. In Ref.~\cite{Pavlov} the first few terms of the
expansion was found for billiards. Here however, the discussion was specific
to quantum billiards and it does not seem easy to generalise it to
potential systems.

The structure of this paper is as follows.
In section~2 we discuss the mechanics of the decomposition of the level
density into the irreducible representations.
Both the density operator and the projection operators are evaluated
using the Wigner representations of the operators so this topic is
briefly reviewed. We discuss discrete groups in section~3
and the continuous group $SO(n)$ in section~4.
All point group elements can be expressed as a multiple reflections,
in $n$ dimensions there are at most $n$ independent reflections.
The reduction of more complicated group elements
to multiple reflections allows for an $\hbar$-expansion of the operators.
For continuous groups, we find that
the semiclassical limit and the limit of small rotation angles are
noncommuting. The results we obtain are found to only be sensible if we further
average over angular momentum.  In section~5 we discuss the H\'{e}non-Heiles
potential which has a $C_{3v}$ discrete symmetry.  We compare the results of
sections 2 and 3 with the actual spectra. By studying specific combinations
of the reduced level densities, we can isolate the contribution from
different group elements. We find that we can also discern the effects of
the shortest periodic orbits on the oscillating density of states, as
discussed in section~6. Section~7 contains some concluding remarks.

\section{The Formalism}

We are interested in the general class of quantum Hamiltonians,
\be \label{qham}
\hat{H}={\hat{\bf p}^2\over 2}+V(\hat{\bf q})
\ee
for which the potential $V(\hat{\bf q})$ is
invariant with respect to the symmetry transformations of a group $G$.
In the present paper, we consider all groups $G$ that are subgroups of
the group of orthogonal transformations in $n$ dimensions, $O(n)$.
The groups $G$ can include improper
rotations, and can be either discrete or continuous.
These groups clearly cover many physical applications, they
include the crystallographic point groups and the symmetric group $S_n$
of $n$ identical particles.

The Hamiltonian commutes with all the group operators of $G$
and the eigenstates of the Hamiltonian can be classified  according to
the irreducible representations (irreps) of the group.
This means that we can decompose the full density of states into
densities of states belonging to each irrep,
\begin{equation}
\rho(E) = \sum_\alpha\rho_\alpha(E),
\end{equation}
where the index $\alpha$ runs over all the irreps of $G$.
Corresponding formulae apply for $\bar{\rho}(E)$ and $\rho_{\rm osc}(E)$.
Each partial density $\rho_\alpha(E)$ may be obtained by
projection,
\begin{equation} \label{rhorep}
\rho_\alpha(E) = \mbox{tr}\{\hat{P}_\alpha\delta(E-\hat{H})\},
\end{equation}
where the operator $\hat{P}_\alpha$ projects onto the representation $\alpha$
and is given by \cite{Ham}
\begin{eqnarray}
\hat{P}_\alpha & = &
\frac{d_\alpha}{|G|}\sum_g\chi_\alpha(g)\hat{U}^{\dagger}_g \label{proj}
\nonumber\\
    & = & \int d\mu(g)\chi_\alpha(g)\hat{U}^{\dagger}_g
         ~ /\int d\mu(g).
\end{eqnarray}
The first expression applies when the group $G$ is discrete; $d_\alpha$
is the dimension of the irrep and $|G|$ is the order of the group.
The sum is over all of the group elements
$g\in G$ and $\chi_\alpha(g)$ is the
character of element $g$ in representation $\alpha$.
The operator $\hat{U}_g$ transforms wavefunctions as
prescribed by the group element $g$.
The second expression applies when $G$ is
continuous, in which case the sum is replaced by the integral over the group
elements with the invariant Haar measure $d\mu(g)$.

In the semiclassical limit, average densities are conveniently described
by phase space integrals through the use of the Wigner representations of the
various operators. The Wigner transform of a generic quantum operator
$\hat{A}$ is defined as
\begin{equation} \label{wigner}
A_W({\bf q},{\bf p}) = \int d{\bf x}~
e^{-i{\bf p}\cdot{\bf x}/\hbar}~
{}~\langle{\bf q}+
\frac{{\bf x}}{2}|\hat{A}|{\bf q}-
\frac{{\bf x}}{2}\rangle.
\end{equation}
This has the appealing property of being a complete representation of the
quantum operator defined in the classical phase space.
The phase space is assumed to be $2n$-dimensional so
${\bf q}$, ${\bf p}$ and $\bf x$
are all $n$-dimensional vectors.  Traces have a simple evaluation in
this representation since
\begin{eqnarray}
\mbox{tr}\{\hat{A}\} & = & \int \frac{d{\bf p}
d{\bf q}}{(2\pi\hbar)^n}
{}~A_W({\bf p},{\bf q}) \label{trwig} \nonumber \\
\mbox{tr}\{\hat{A}\hat{B}\} & = & \int \frac{d{\bf p}
d{\bf q}}{(2\pi\hbar)^n}
{}~A_W({\bf q},{\bf p})B_W({\bf q},
{\bf p}).
\end{eqnarray}
These equations do not generalise to more than two noncommuting
operators as can be seen
from the fact that the ordering of the operators in the traces is important
while the ordering in the integrals is irrelevant.
Instead we use the product rule,
\begin{equation} \label{comb}
\left(\hat{A}\hat{B}\right)_W =
A_W e^{i\hbar\stackrel{\leftrightarrow}{\Lambda}/2}B_W
\end{equation}
where the Moyal operator $\stackrel{\leftrightarrow}{\Lambda}$ is given by
\begin{equation} \label{moyal}
\stackrel{\leftrightarrow}{\Lambda} = \stackrel{\leftarrow}{\nabla}_q \cdot
\stackrel{\rightarrow}{\nabla}_p - \stackrel{\leftarrow}{\nabla}_p \cdot
\stackrel{\rightarrow}{\nabla}_q
\end{equation}
and the arrows indicate which way the derivatives should be applied.
Eqs.~(\ref{trwig}), (\ref{comb}) and (\ref{moyal}) allow us in principle
to evaluate the trace of any operator of the form
$\hat{A}\hat{B}\hat{C}\cdots$ given the Wigner transforms
of the individual operators.
A final property of the Wigner transform which
we will use is that Hermitian conjugation transforms into complex
conjugation, $(\hat{A}^\dagger)_W= (A_W)^*$. In particular,
this implies that the Wigner transform of any Hermitian operator is real.

Using the definition (\ref{wigner}) we find that the Wigner transform
of the quantum Hamiltonian (\ref{qham}) is simply the classical
Hamiltonian
\begin{equation} \label{wighamm}
H_W({\bf q},{\bf p}) =
\frac{{\bf p}^2}{2} + V({\bf q}).
\end{equation}
To evaluate the partial densities of Eq.~(\ref{rhorep}) we need the
Wigner transform of the density operator $\delta(E-\hat{H})_W$.
However, it is not true that $\delta(E-\hat{H})_W$ equals $\delta(E-H_W)$. The
difference between these expressions gives the higher order contributions to
the full density of states.  These have been worked out in a particularly
elegant manner in Ref.~\cite{BTU}.
The density operator can be written as the inverse Laplace transform,
\be
\delta(E-\hat{H})_W ={1\over 2\pi i} \int_{-i\infty}^{+i\infty}
d\beta~\left( e^{-\beta \hat{H}}\right)_W e^{\beta E},
\ee
where the Wigner transform of the evolution operator
is expanded in $\hbar$,
\be
\left( e^{-\beta \hat{H}}\right)_W = e^{-\beta H_W}\sum_{m=0}^\infty
{1\over (2m)!} \left( {-\hbar^2\over 4}\right)^m A_m(\beta ) .
\ee
The coefficients $A_m$ are simple polynomials in $\beta$ and obey the
recursion relation,
\be
-{d\over d\beta}A_m=e^{\beta H_W}
\sum_{k=0}^{m-1}{2m\choose 2k}
\left[H_W \stackrel{\leftrightarrow}{\Lambda}^{(2m-2k)}e^{-\beta H_W}A_m
\right] ;~~A_0=1,
\ee
from which one may easily generate the series.
With the Hamiltonian (3), the first few terms  of the density operator are
\begin{eqnarray}
\delta(E-\hat{H})_W & = & \delta(E-H_W) - \frac{\hbar^2}{8}
{\bf \nabla}^2
V\frac{d^2}{dE^2}\delta(E-H_W) \label{deltwig} \nonumber \\
& & + \frac{\hbar^2}{24}\left(({\bf p}\cdot
{\bf \nabla})^2V +({\bf \nabla}V)^2\right)
\frac{d^3}{dE^3}\delta(E-H_W) + O(\hbar^4),
\end{eqnarray}
a result already obtained by Wigner in 1932 \cite{Wigner}.
An alternative approach, which yields the same expansion, was developed in
Ref.~\cite{vorgramm}.
In calculating the density of states, the derivatives with respect to
energy can be taken outside the phase space integral.

The semiclassical expansion of the density operator only contains even powers
of $\hbar$. This is in contrast to the expansion for quantum billiards,
which in general contains all integer powers of $\hbar$, the first odd
term arising from the Dirichlet or Neumann boundary conditions.
In general, quantum billiards require a separate treatment; the
semiclassical expansions of the Wigner transforms above do not converge
when the billiard boundaries are approximated by steep but smooth potential
walls. We will find that the projection operators (\ref{proj}) also
contain odd powers of $\hbar$ in their Wigner expansions.
In simple cases, the various irreps may
be realized by internal hard wall boundaries thus explaining the
odd powers. However, the semiclassical expansion as stated does not allow
for singular boundaries.
Instead, we will directly evaluate the Wigner transforms of the projection
operators $P_\alpha$ or equivalently of the transformation
operators $\hat{U}^\dagger_g$.

Eq.~(\ref{deltwig}) appears to contain only the smooth part of the
density of states. The derivation is certainly exact so we should expect the
trace  of the density operator to contain the oscillating term in addition to
the average term.  The resolution of this apparent paradox
lies in the concept of resurgence \cite{Berry}.
The Weyl series is an asymptotic expansion in $\hbar$ and
information about the oscillating density of states is contained in the very
high order terms.  In that sense, there is no loss of information in
Eq.~(\ref{deltwig}),
only when we truncate the series we disregard the oscillations in the
density of states.
The truncated Weyl series gives the average density of states. It
follows, that the separation between the average and oscillating terms of
the density of states is somewhat arbitrary. In principle, we could keep all
terms in the expansion until the terms start to grow. In practise, we
are satisfied with a few terms, e.g. all terms up to order
$O(\hbar^{n+1})$ in Eq.~(\ref{deltwig}) or equivalently, all terms up to order
$O(\hbar)$ in the density of states.

\section{Discrete Groups}

The point groups in $n$ dimensions are subgroups of the group of
orthogonal transformations $O(n)$ and the group elements can be represented
by $n\times n$ orthogonal matrices acting on the coordinates.
The eigenvalues of an orthogonal matrix $R\in O(n)$ have unit norm and
either come in complex conjugate pairs, $\lambda =\exp (\pm i\phi)$ or
are real, $\lambda = \pm 1$.
An eigenvalue $\lambda =-1$ corresponds to a reflection in a hyperplane
of dimension $n-1$ while the complex eigenvalues $\exp (\pm i\phi)$
correspond to a rotation by the real angle $\phi$ in a hyperplane of
dimension $2$. It follows that by an orthogonal transformation, $R$ can be
brought into a block-diagonal form,
\be \label{decomp1}
R_D = \pmatrix{
r_1&0&\ldots&0\cr
0&r_2&\ldots&0\cr
\vdots&\vdots&\ddots&\vdots\cr
0&0&\ldots&r_k\cr}
\ee
where the blocks are either one-dimensional $r_i=\pm 1$ or
two-dimensional of the form
\be
r_i = \pmatrix{
\cos\phi_i & -\sin\phi_i \cr
\sin\phi_i & \cos\phi_i \cr} .
\ee
Thus any orthogonal transformation in $n$ dimensions can be
constructed out of $k$ successive, independent transformations,
\be \label{mmult}
R = R_k \cdots R_2 R_1
\ee
where each transformation $R_i$ is either a reflection or a rotation.
The transformations $R_i$ commute since they act in orthogonal hyperplanes
and the decomposition (\ref{mmult}) is unique up to an irrelevant ordering
of the $k$ transformations.
In three dimensions we infer that there are only four types of group
elements, the identity $I$, a simple reflection $R_\sigma$, a proper
rotation $R_\phi$ and an improper rotation $R_\sigma R_\phi$.
For the improper rotation, $R_\sigma$ is a reflection through the
rotational plane.

Corresponding to each coordinate transformation $R_i$, there is a quantum
operator $\hat{U_i}=\hat{U}(R_i)$ which operates on the Hilbert space. These
operators have the same multiplication properties as (\ref{mmult}) and
are defined by
\be
\hat{U_i}\psi({\bf r}) = \psi(R_i^{-1}{\bf r}).
\ee
We can now evaluate the density of states belonging to the
irrep $\alpha$ of a discrete symmetry group $G$.
{}From Eqs.~(\ref{rhorep}), (\ref{proj}) and (\ref{trwig}) we have
\begin{equation} \label{rhored}
\rho_\alpha(E) = \frac{d_\alpha}{|G|}\sum_g\chi_\alpha(g)
\int \frac{d{\bf p}d{\bf q}}{(2\pi\hbar)^n}
{}~(\hat{U}_g^\dagger)_W \delta(E-\hat{H})_W.
\end{equation}
The semiclassical expansion of $\delta(E-H)_W$ is given by
Eq.~(\ref{deltwig}). The Wigner transform $(\hat{U}_g)_W$ is most easily
evaluated in a coordinate
basis where the matrix representation of the group element $g$
attains the block-diagonal form (\ref{decomp1}).
By direct application of the definition, Eq.~(\ref{wigner}), it is clear
that with the decomposition (\ref{mmult})
the Wigner transform $(\hat{U}_g)_W$ also factorises as
\be \label{decomp2}
(\hat{U}_g)_W = (\hat{U}_k)_W \cdots (\hat{U}_2)_W (\hat{U}_1)_W.
\ee
Each $(\hat{U}_i)_W({\bf q},{\bf p})$ is a function of only those
coordinates and momenta that are not left invariant by the transformation.
The factorization (\ref{decomp2}) could also have been obtained by repeated
applications of the product rule (\ref{comb}), using the orthogonality
of the invariant hyperplanes of the transformations $R_i$.

It remains to evaluate the Wigner transform of the basic operators
$\hat{U}^\dagger_i$. The identity element $I$ is trivial since
\be
(\hat{U}_I)_W = 1,
\ee
corresponding to an empty product (\ref{decomp2}).
{}From Eq.~(\ref{wigner}) it also follows that the Wigner transform of
a reflection through the $q_1=0$ hyperplane is given by
\begin{eqnarray}
(\hat{U}_\sigma )_W & = & \int dx_1 e^{-ip_1x_1/\hbar}
\langle q_1+\frac{x_1}{2}|-q_1+\frac{x_1}{2}\rangle \label{ref}\\
                                & = & \pi\hbar\delta(q_1)\delta(p_1). \nonumber
\end{eqnarray}
The quantum operator corresponding to a rotation by an angle $\phi$
in the $q_1 - q_2$ plane is given by
$\hat{U}_\phi=\exp{(-i\phi\hat{L}_3/\hbar)}$.
Its adjoint has the Wigner transform,
\be \label{rottemp}
(\hat{U}^\dagger_\phi)_W=
\frac{e^{2i\tan (\phi /2)(q_1p_2-q_2p_1)/\hbar}}{\cos^2(\phi /2)}.
\ee
The Wigner transforms of the identity and reflection operators
automatically lend themselves to a semiclassical expansion.
The rotation operator on the other hand,
poses a problem as its Wigner transform (\ref{rottemp}) contains an essential
singularity in $\hbar =0$.
This may however, be circumvented in the following manner.

Each two-dimensional rotation by an angle $\phi$ can be constructed from
two consecutive reflections about two axes (located in the
two-dimensional hyperplane) that intersect at an angle $\phi /2$.
Two such reflections do not commute, the opposite ordering
would amount to an inverse rotation.
{}From the product rule (\ref{comb}) and the Wigner transform of a reflection
operator (\ref{ref}) we may write
\be \label{rot}
(\hat{U}^\dagger_\phi )_W= (\pi\hbar )^2 \delta(q'_1)\delta(p'_1)
e^{-i\hbar\stackrel{\leftrightarrow}{\Lambda}/2}
\delta(q_1)\delta(p_1)
\ee
where $q_1=0$ and $q'_1=0$ denote the non-orthogonal invariant hyperplanes
of the two reflections.
The essential singularity is now only implicit through the derivatives
of the distribution functions $\delta(q)$, $\delta(p)$, and we shall
simply expand the expression (\ref{rot}) in $\hbar$.
Since the $q'_1=0$ hyperplane is rotated from the $q_1=0$ hyperplane by
the angle $\phi /2$ we obtain,
\bea
(\hat{U}^\dagger_\phi )_W & = & (\pi\hbar )^2
\delta ( \cos({\phi\over 2})q_1+\sin({\phi\over 2})q_2 )
\delta ( \cos({\phi\over 2})p_1+\sin({\phi\over 2})p_2 )
e^{-i\hbar\stackrel{\leftrightarrow}{\Lambda}/2}
\delta(q_1)\delta(p_1) \nonumber \\
& = & {(\pi\hbar )^2 \over \sin^2(\phi/2)} \delta(q_2)\delta(p_2)
\nonumber \\
&&\times \sum_{m=0}^\infty {1\over m!}
\left( {-i\hbar\over 2} \cot({\phi\over 2})\right)^m~
\left(\ddl{q_2}\ddr{p_1}-\ddl{p_2}\ddr{q_1} \right)^m
\delta(q_1)\delta(p_1) . \label{junk}
\eea

To leading order in $\hbar$ we have derived the simple result
\be \label{junk2}
(\hat{U}^\dagger_\phi )_W \approx {(\pi\hbar )^2 \over \sin^2(\phi/2)}
\delta(q_1)\delta(p_1)\delta(q_2)\delta(p_2)
\ee
also valid for billiard systems \cite{Pavlov}.
The higher order terms containing the partial derivatives may evaluated
when we perform the phase-space integral of the reduced
density of states, Eq.~(\ref{rhored}).
That is, integrating by parts we obtain for the ``rotational density'',
\bea
\rho_\phi &\equiv&{\rm tr}\{\hat{U}^\dagger_\phi\delta(E-\hat{H})\}
=\int {d{\bf p}d{\bf q}\over (2\pi\hbar )^n}
{}~(\hat{U}_\phi^\dagger )_W \delta(E-\hat{H})_W  \nonumber \\
&=& {(\pi\hbar)^2 \over \sin^2(\phi/2)}
\int {d{\bf p}d{\bf q}\over (2\pi\hbar )^n}
{}~\delta(q_1)\delta(p_1)\delta(q_2)\delta(p_2)  \label{messy} \nonumber \\
&& \times \sum_{m=0}^\infty {1\over m!}
\left( {i\hbar\over 2} \cot(\phi/2)\right)^m~\left( \ddd{q_1}{p_2}
- \ddd{q_2}{p_1} \right)^m \delta(E-\hat{H})_W.
\eea
All the derivatives now act on the Wigner transform of the density
operator, $\delta(E-\hat{H})_W$. With the expansion (\ref{deltwig}) of the
density operator, it is straightforward,
if tedious, to write down the integrals of Eq.~(\ref{messy}) in closed form.
We do not do so here but only discuss the leading order contributions
to each partial density of states $\rho_\alpha$.

The identity operator is present in any symmetry group $G$.
In the semiclassical expansion, this gives the leading order term for all
representations.
Recalling that the character of the identity element is simply the dimension of
the representation we get as a first approximation
\begin{equation} \label{crude}
\rho_\alpha(E) \approx \frac{d_\alpha^2}{|G|}\rho(E),
\end{equation}
where $\rho(E)$ is the total density of states. This was also noted in
Ref.~\cite{Pavlov} where the author argued that this expression holds
equally for potential systems and billiard problems.
The next to leading order term is given by single reflection operators if they
are present  in the group. Rotation operators contribute to the second order
corrections as seen in Eq.~(\ref{junk}). As mentioned in the last section,
reflection symmetries lead to the odd powers of $\hbar$ which are absent in
the full density of states.

In the general case, more complicated group elements in the sense of the
factorisation (\ref{decomp2}) contribute to higher orders in $\hbar$.
That is, for each group element $g$, the lowest order in $\hbar$ of the
expansion of $(\hat{U}_g)_W$ is given by the number of independent reflections
contained in $g$, a rotation counting as two reflections.
Therefore, the power of $\hbar$ with which $\hat{U}_g$ contributes is
the codimension of the set of points which is invariant with respect to
the group element $g$.
Loosely speaking, it is the
invariant points which contribute to the zero length pseudo-orbits from which
the Weyl term can be evaluated, as argued in Ref.~\cite{Pavlov}.
For example, the points which are invariant under a finite rotation are
those along the axis of rotation and comprise a set of codimension two.
Indeed, the corresponding operator contributes with two added powers of $\hbar$
relative to the identity operator.
Note in this case, that the limit of a small angle $\phi$,
the contribution diverges.  This is because all points become
approximately invariant in this limit.  The difficulty is that the limits
$\hbar\rightarrow 0$ and $\phi\rightarrow 0$ do not commute; we discuss this
problem in the next section.


To conclude this section we give the explicit expansions for a
two-dimensional potential system, keeping only terms with non-positive
powers of $\hbar$. The remainder of the smooth density of states then
vanishes in the limit of $\hbar\rightarrow 0$.
Corresponding formula for higher dimensional systems may easily be obtained
with the help of the factorization (\ref{rhored}).
It is convenient to work with the integrated density of states
defined by
\begin{equation} \label{intdef}
N_\alpha(E) = \int_{-\infty}^EdE'\rho_\alpha(E')
\end{equation}
which is the number of states in representation $\alpha$ with energy less than
$E$. We then find from Eq.~(\ref{rhored})
\begin{equation} \label{intred}
N_\alpha(E) = \frac{d_\alpha}{|G|}\sum_g\chi_\alpha(g)N_g(E),
\end{equation}
where we have defined
\begin{equation} \label{intgrp}
N_g(E) \equiv \int_{-\infty}^EdE'\int
\frac{d{\bf p}d{\bf q}}{(2\pi\hbar)^n}~
(\hat{U}^\dagger_g)_W
\delta(E-\hat{H})_W .
\end{equation}
The quantity $N_g(E)$ is the contribution to the various integrated
densities of states from group element $g$.
We obtain up to order $\hbar$,
%
\begin{eqnarray} \label{disres}
N_I(E) & = & \int {d{\bf p}d{\bf q}\over(2\pi\hbar)^2}
\left [ ~\Theta(E-H_W) - \frac{\hbar^2}{8}
\frac{d^2}{dE^2}\Theta(E-H_W) {\bf \nabla}^2
V({\bf q}) \right. \nonumber \\
&& \left. ~~~~~~~~ + \frac{\hbar^2}{24} \frac{d^3}{dE^3}\Theta(E-H_W)
\left(({\bf p}\cdot{\bf \nabla})^2V +
({\bf \nabla}V)^2\right) \right ] +
O(\hbar^2) \nonumber \\
N_\sigma(E) & = & {1\over 2} \int {dq_2dp_2\over 2\pi\hbar }
{}~\Theta(E-H_W)|_{q_1=p_1=0} + O(\hbar) \nonumber \\
N_\phi(E) & = & \frac{1}{4\sin^2(\phi/2)}
{}~\Theta(E-H_W)|_{{\bf q}={\bf p}=0}  + O(\hbar^2),
\end{eqnarray}
where $\Theta(x)$ is the Heaviside step function and the
reflection $\sigma$ is defined to be in the $q_1=0$ axis.

\section{Continuous Groups }

Much of the discussion of the discrete point groups also applies to the
continuous point groups.
We begin by studying the abelian group $SO(2)$.
However, as we argued in the last section, we may generalize the
results to the group of rotations in $n$ dimensions, $SO(n)$, by use of
decompositions of the form (\ref{mmult}).
The $SO(2)$ group is also relevant to integrable Hamiltonians
where the $SO(2)$ symmetry is manifest in the action-angle variables since
the Hamiltonian does not depend on the angles.

We define the rotation to be in the $q_1-q_2$ plane and denote by
$\hat{L}_3$ the corresponding angular momentum operator.
The irreps are then labelled by the angular momentum eigenvalues $m$
and the projection operator is given by Eq.~(\ref{proj}),
\begin{equation} \label{so2proj}
\hat{P}_m = \frac{1}{2\pi}\int d\theta
{}~e^{-im\theta}\hat{U}^\dagger_\theta.
\end{equation}
with the rotational operator
\be
\hat{U}^\dagger_\theta = \exp{(i\theta\hat{L}_3/\hbar)}.
\ee
{}From the form of the rotational operator it is clear that for
continuous groups the semiclassical limit,
$\hbar \rightarrow 0$, does not commute with the small angle limit,
$\theta \rightarrow 0$.
This is also obvious from the expansion of the ``rotational density'',
Eq.~(\ref{junk2}), where all terms diverge in the limit of small angles.
Instead, we need to project {\it before} we expand semiclassically.
The Wigner transform of the rotation operator is given by Eq.~(\ref{rottemp}),
\begin{equation} \label{so2wig}
\left(\hat{U}^\dagger_\theta\right)_W = \frac{e^{2i\tan{(\theta/2)}L_3/\hbar}}
{\cos^2{(\theta/2)}},
\end{equation}
where $L_3$ is the Wigner representation of $\hat{L}_3$ and equals the
classical angular momentum $q_1p_2-q_2p_1$.
The density of states of angular momentum $\langle\hat{L}_3\rangle=m$
then reads
\begin{equation} \label{rhom1}
\rho_m(E) = \frac{1}{2\pi}\int
\frac{d{\bf p}d{\bf q}}{(2\pi\hbar)^n}
\int d\theta
{}~\frac{e^{-im\theta}}{\cos^2(\theta/2)}
e^{2i\tan{(\theta/2)}L_3/\hbar}
\delta\left(E-\hat{H}\right)_W.
\end{equation}

This expression may be brought into a more convenient form for
semiclassical expansion by substituting the
variable $z=\tan{(\theta/2)}/\hbar$ to obtain
\begin{equation} \label{rhom2}
\rho_m(E) = \int\frac{d{\bf p}d{\bf q}}
{(2\pi\hbar)^n}
{}~\frac{\hbar}{\pi} \int_{-\infty}^{\infty} dz
\left(\frac{1-i\hbar z}{1+i\hbar z}\right)^m
e^{2izL_3}~\delta(E-\hat{H})_W.
\end{equation}
We separately consider $m=0$ and $m\neq 0$.
For $m =0$ the integral over $z$ is trivial, so the density of
zero angular momentum states becomes,
\begin{equation} \label{rhomis0}
\rho_0(E) =
\int \frac{d{\bf p}d{\bf q}}
{(2\pi\hbar)^n}~\hbar\delta(L_3)\delta(E-\hat{H})_W.
\end{equation}
This result has the simple interpretation that the $m=0$ states correspond
directly to the $L_3=0$ region of phase space. If we are only interested in
such states we can ignore the rest of phase space and quantise directly on the
$L_3=0$ manifold.
The result may be slightly surprising as it seems to neglect quantal
fluctuations around that manifold.

For non-zero values of $m$ the integral can also be evaluated exactly
but it will not lead to transparent results. Instead we expand the
integrand in a formal series
\begin{eqnarray}
\left(\frac{1-iz\hbar}{1+iz\hbar}\right)^m & = & \exp\left(-2imz\hbar
\sum_{k=0}^\infty\frac{(-iz\hbar )^{2k}}{2k+1}\right) \label{fracsim}
\nonumber \\
& = & e^{-2miz\hbar}\left(1 + \frac{2m}{3}(-iz\hbar )^3 +
\cdots\right),
\end{eqnarray}
from which we obtain the desired semiclassical expansion of the
density of states belonging to the irrep $m$,
\begin{equation} \label{rhoisnt0}
\rho_m(E) =
\int {d{\bf p}d{\bf q}\over (2\pi\hbar )^n}
\left(\hbar\delta(L_3-m\hbar)
+\frac{m}{12}\hbar^4\delta^{(3)}
(L_3-m\hbar) + \cdots\right)\delta(E-\hat{H})_W.
\end{equation}
As for $m=0$ we have an identification between states with a given value of
$m$ and the corresponding $L_3=m\hbar$ region of phase space. However, there
are corrections in the form of derivatives of delta functions. It is not enough
to consider just the $L_3=m\hbar$ manifold; we must also understand its
neighbourhood in order to evaluate the derivatives. In addition, it should be
noted that the expansion in Eq.~(\ref{fracsim}) is convergent only if
$|i\hbar z|<1$ so strictly speaking it is incorrect to integrate $z$ with this
expansion all the
way to infinity. However, the error in doing so is small since the integral
(\ref{rhom2}) is highly oscillatory for large $z$.  Furthermore, the
oscillations are washed out if we further average over a range of $m$ values.
Therefore, we interpret Eq.~(\ref{rhoisnt0}) as the average density of
states in both energy and angular momentum.  Both of these averages are over
ranges which are quantum mechanically large but classically small. Therefore
this result only holds for $m\gg 1$. This may be expected since it simply says
that semiclassics works in the limit of large quantum numbers.

The lowest order term of the expansion (\ref{rhoisnt0}) was obtained in
Ref.~\cite{sccthes} by integrating over the short orbits
in a symmetry reduced Green function.
The result here is slightly more general
since we have shown, in principle, how to obtain the higher order terms. In
addition, we have shown that the case of $m=0$ is special since the na\"{\i}ve
expectation is exact in this case.

The generalisation to higher dimensional systems is now straightforward.
If the Hamiltonian has the symmetry of $k$ independent $SO(2)$ groups,
the eigenstates can be labelled by $k$ quantum numbers.
To leading order, the partial densities are
\begin{equation} \label{stephres}
\rho_{\bf m}(E) =
\int\frac{ d{\bf p}d{\bf q}}{(2\pi\hbar)^n}
{}~\hbar^k\delta^k({\bf L}-\hbar{\bf m})
\delta(E-\hat{H})_W.
\end{equation}
where
${\bf m}$ is the $k$ dimensional vector of integers which labels the
representations of the $k$ groups and $\bf L$ is the vector of
the corresponding classical generators.

Rotational invariance in three dimensions is described by the symmetry
group $SO(3)$. The irreps are labelled by the angular momentum $l$ and
are $(2l+1)$-fold degenerate. The characters of each irrep $l$ is
$\chi_l(\theta) = \sin ((l+{1\over 2})\theta)/\sin({\theta\over 2})$
so that the projection operator becomes \cite{Ham}
\be \label{so3proj}
\hat{P}_l={2l+1\over 4\pi}\int d\Omega \int d\theta \sin^2({\theta\over 2})
{\sin ((l+{1\over 2})\theta)\over\sin({\theta\over 2})}
e^{i\theta\hat{L}_\Omega /\hbar}.
\ee
Here $\Omega$ gives the direction of the rotation axis and
$\hat{L}_\Omega(\theta)$ is the corresponding angular momentum operator.
The Wigner transform of the rotation operator is the same as in the $SO(2)$
calculation and is given by Eq.~(\ref{so2wig})
\be \label{so3wig}
\left(e^{i\hat{L}_\Omega \theta/\hbar}\right)_W =
\frac{e^{2i\tan{(\theta/2)}L_\Omega /\hbar}} {\cos^2{(\theta/2)}},
\ee
where $L_\Omega$ now denote the value of classical angular momentum along
the rotation axis.
Substituting this for the rotation operator in Eq.~(\ref{so3proj})
and integrating
over $\Omega$ we obtain the Wigner transform of the projection operator,
\be
(\hat{P}_l)_W={2l+1\over 2\pi} \int_{-\pi}^\pi d\theta
{}~{\sin ((l+{1\over 2})\theta)\over\cos({\theta\over 2})}
{}~{\hbar\over L} \sin\left(2\tan({\theta L\over 2\hbar})\right) .
\ee
Here $L$ is the length of the angular momentum vector.
As in the two-dimensional case, we substitute for $\theta$ the variable
$z=\tan{(\theta/2)}/\hbar$ and expand in $\hbar$,
\begin{eqnarray}
\theta = 2\tan^{-1}(z\hbar)
&=& 2\sum_{k=0}^\infty {(-1)^k\over 2k+1}(z\hbar )^{2k+1} \nonumber\\
&=& 2\left(z\hbar - {1\over 3}(z\hbar )^3 +
\cdots\right).
\end{eqnarray}
To first order in $\hbar$ the Wigner transform of the projection
operator takes the simple form,
\bea
(P_l)_W &=& (2l+1)
{\hbar\over 2L} \delta(l+{1\over 2} -{L\over \hbar}) \nonumber \\
&=& (2l+1) \hbar^2  \delta(L^2 -(l+{1\over 2} )^2\hbar^2 )
\eea
and the density of states of angular momentum $l$ is given by
\be
\rho_l(E) =
(2l+1)\int\frac{d{\bf p}d{\bf q}}
{(2\pi\hbar)^3}~\hbar^2
\delta\left(L^2 -(l+{1\over 2})^2\hbar^2\right)\delta(E-\hat{H})_W.
\ee
For any $L$ we have the result that we must evaluate quantities on the manifold
$L=(l+1/2)\hbar$. As before, the corrections to this expression
involve derivatives evaluated on
the manifold. Note that unlike in the case of $SO(2)$,
there is nothing special about the $L=0$ states.

For higher dimensional groups $SO(n)$, each orthogonal
transformation decomposes as in Eq.~(\ref{mmult}) into a sequence of
mutually orthogonal rotations $R_i=\exp(i\theta_i\hat{L}_{i})$
and we must determine the Wigner transform of a product of the form
$\exp(i\theta_1\hat{L}_{1})\exp(i\theta_2\hat{L}_{2})\cdots$.
This is a trivial extension of what we have done for $SO(2)$ so in principle
we have worked out the semiclassical expansion. Of course, for each group
$SO(n)$, one must include the characters specific to that group.

\section{A Numerical Example}

We shall verify the results of section 3 by evaluating the partial
density of states in a simple symmetric potential system in two dimensions.
Symmetry transformations in higher dimensional systems may, with respect to
their semiclassical expansions, be reduced to a sequence of reflections and
rotations, already present in two dimensions.
We use the H\'{e}non-Heiles potential \cite{hh}
\begin{eqnarray} \label{hhpot}
V(x,y) & = & \frac{1}{2}(x^2+y^2) + x^2y-\frac{1}{3}y^3
\nonumber\\
       & = & \frac{1}{2}r^2 + \frac{1}{3} r^3 \sin{3\theta},
\end{eqnarray}
first introduced as a model of celestial mechanics.
The potential can be infinitely negative but if $E<1/6$, motion near the origin
remains bounded and for small energies resembles that of
a two dimensional harmonic oscillator.

The potential is symmetric with respect to rotations by an angle
$\phi=\pm 2\pi/3$ and with respect to
reflections through three symmetry axes. Together with the identity element,
these comprise the three classes of the six member symmetry group $C_{3v}$.
Note that
this is isomorphic to the permutation group $S_3$ which arises when we consider
three identical particles interacting in one dimension.
These groups have three irreducible representations. Two are one dimensional
and are labelled $A_1$ and $A_2$.
The third is two dimensional and is labelled $E$.
The character table is shown in Table~1.

Using Eq.~(\ref{intred}) we find
\begin{eqnarray}
N_{A_1}(E) & = & \frac{1}{6} \left[N_I(E) + 2N_\phi(E) + 3N_\sigma(E) \right]
\nonumber \\
N_{A_2}(E) & = & \frac{1}{6} \left[N_I(E) + 2N_\phi(E) - 3N_\sigma(E) \right]
\nonumber \\
N_E(E) & = & \frac{2}{6} \left[2N_I(E) - 2N_\phi(E)\right] \label{aberep}
\end{eqnarray}
We have used the fact that the contribution from each member of a class is
identical.  Note that the leading order contribution has the densities in the
ratios 1:1:4 as we expect from Eq.~(\ref{crude}).  We could calculate
these three quantities separately in order to compare with the numerics.
However, it is more instructive to isolate the contribution
$N_g(E)$ from each symmetry class by inverting Eq.~(\ref{aberep}) as follows,
\begin{eqnarray} \label{nr}
N_I(E) & = & N_{A_1}(E) + N_{A_2}(E) + N_E(E)\nonumber \\
N_\sigma(E)  & = & N_{A_1}(E) - N_{A_2}(E)\nonumber \\
N_\phi(E)    & = & N_{A_1}(E) + N_{A_2}(E) - N_E(E)/2.
\end{eqnarray}
The first of these is just the total density of states and we do not
consider it any further.
The semiclassical expansions of the other two terms are given by equation
(\ref{disres}).
For the H\'{e}non-Heiles potential they are to leading order
\begin{eqnarray}
N_\sigma(E)    & = & \int\frac{dydp_y}{4\pi\hbar}
{}~\Theta\left(E-(\frac{1}{2}p_y^2 + \frac{1}{2}y^2 -
\frac{1}{3}y^3)\right) \label{res1} \\
N_\phi(E) & = & \frac{1}{3} . \label{res2}
\end{eqnarray}
The reflection term $N_\sigma(E)$ has to be be evaluated numerically.
Note that the rotation function $N_\phi(E)$ has no
corrections at finite energy; all correction terms are of the form of
derivatives of the stepfunction $\Theta(E)$, which vanish
identically for $E>0$.

We have calculated the exact quantum spectrum for $\hbar=0.004$ and evaluated
the smooth density of states by convolving the exact density of states of each
representation with a Gaussian of width $w$ so that
\bea \label{rhosmooth}
\tilde{\rho}_\alpha(E) &=&
\frac{1}{\sqrt{2\pi w^2}}\int_{-\infty}^{\infty}dE'
{}~e^{-(E-E')^2 /2w^2} \rho_\alpha (E') \nonumber \\
&=& \sum_n \frac{1}{\sqrt{2\pi w^2}}
\exp\left( -\,{(E-E_{n,\alpha})^2\over 2w^2}\right) ,
\eea
where $E_{n,\alpha}$ is the energy of the n'th state in representation
$\alpha$.
We use the notation $\tilde{\rho}$ for the smoothed density of states in
order to distinguish it from the semiclassical average $\bar{\rho}$.
For consistency, we must smooth both $\bar{\rho}$
and the exact density of states; these smoothed quantities should agree
when the smoothing width $w$ is large.

The results for the reasonably large value of $w=.01$ are shown if Fig.~1.
The solid lines are evaluated from the exact spectrum while the dashed
lines are the results shown in equations (\ref{res1}) and (\ref{res2}).
In both cases the agreement is very good, in fact for $N_\sigma(E)$ the dashed
lines
lies directly on the solid line and can not be discerned.  Also note that the
behaviour at small energy is modified by the smoothing; for $N_\phi(E)$ the
curves  begin at the value $1/6$ rather than $1/3$ while for $N_\sigma(E)$ the
curves do not begin at the origin as they would for the unsmoothed case.
There is also discrepancy at the high end of the spectrum due to the fact that
the calculation of the quantum energies becomes unreliable near the threshold
of $E=1/6$.  This affects $N_\phi$ more than $N_\sigma(E)$ since it relies on
a higher order of cancellation (note the scale) and is therefore more
sensitive to inaccuracies.

\section{Effect of Periodic Orbits}

As shown in Fig.~1, the discussion of the previous sections accurately gives
the smooth level density for each representation.  To get this agreement it
was necessary to convolve the delta functions of the exact spectrum with a
Gaussian in order to smooth out the spectrum. However, as the width
$w$ of the Gaussian is decreased, we should expect first to see the
periodic orbit structure in the spectrum and
in the limit $w\rightarrow 0$ to recover the exact spectrum.
Here we consider only a relatively large smoothing width, so that
only the shortest periodic orbits are apparent.

We will discuss the periodic orbit structure not in
the full configuration space but rather in one of its fundamental
domains which tessellate the full configuration space.
Dynamics in the fundamental
domain are defined as the dynamics in the full domain mapped onto
the fundamental domain.
In particular, periodic orbits of the full dynamics always map onto
periodic orbits in the fundamental domain.
Conversely, periodic orbits in the fundamental domain correspond to
fractions of periodic orbits in the full domain. These {\it fractional
periodic orbits} have endpoints which
are related by one of the group elements. We can thus associate a group element
to each periodic orbit in the fundamental domain.
However, the group element depends on the choice of fundamental domain
and is only specified up to its class.
Orbits along the boundary of the fundamental domain are exceptional since
group elements from more than one class may connect their endpoints
in the full domain. The treatment of boundary orbits is intricate and, since
we do not consider them in what follows, we forego an involved discussion.

For interior orbits (i.e. non-boundary) the oscillating term of the
partial densities is given by
$\rho_{\alpha,{\rm osc}}=-\mbox{Im}\{g_\alpha\}/\pi$ where \cite{Lau}
\begin{equation} \label{g(E)}
g_\alpha (E) = \frac{d_\alpha}{i\hbar}
\sum_{j\in \rm ppo} \bar{T}_j
\sum_{r=1}^\infty\sum_{k=0}^\infty\chi_\alpha (g_j^r)
\exp{\left(ir\left[\bar{S}_j/\hbar+i(k+1/2)
(\bar{u}_j+i\bar{v}_j)\right]\right)}.
\end{equation}
The index $j$ labels all the primitive
periodic orbits in the fundamental domain and
the $g_j$ is the group element associated with orbit $j$.
The index $r$ counts
the repetitions of the primitive orbits.
$T$, $S$, $u$ and $v$ denote the period, action,
stability exponent and angle respectively, and the bar denote the values
when we consider the dynamics in the fundamental domain.
This is a direct generalisation of the Gutzwiller-Voros formula
\cite{gutzvor} in the presence of a discrete symmetry.

Similarly to Eqs.~(\ref{nr}), we may define the ``class greens
functions'',
\bea \label{nr1}
g_I(E) & = & g_{A_1}(E) + g_{A_2}(E)+g_E(E) \nonumber \\
g_\sigma(E) & = & g_{A_1}(E) - g_{A_2}(E)\nonumber \\
g_\phi(E)    & = & g_{A_1}(E) + g_{A_2}(E) - g_E(E)/2.
\eea
which give the oscillatory contributions to the class densities.

Like any bounded Hamiltonian system, the H\'{e}non-Heiles potential has an
infinity of periodic orbits. However three stand out as special because their
periods are relatively short.
In Fig.~2 we show the configuration space with the three periodic orbits and
the fundamental domain given by the bottom left triangle.
In the figure, we also show the continuation of the periodic orbits in the full
domain as dashed lines.
Following Ref.~\cite{badback} we label the three periodic orbits $A$, $B$
and $C$.
In the fundamental domain, orbit $B$ begins at the short side, reflects off
the hypotenuse to the long side where it turns around and
retraces itself. Its period in the fundamental domain is half that of the full
domain and we associate it with the class of reflections.  Its double
repetition has a full period and we associate it with the identity class. Orbit
$C$ starts at the short side, goes to the hypotenuse and is reflected back. It
has one third
the period of the full orbit and is associated with the rotation class.
Its double repetition is also associated with this class but its
triple repetition has the full period and is associated with the identity
class.
Orbit A runs down the short side to the centre and then up the hypotenuse.
It then turns around and retraces itself so its period in the fundamental
domain is the same as in the full domain.
Since it is a boundary orbit it is associated with both the identity and
the reflection classes.

In the full domain, all three orbits have a period close to $2\pi$,
the value they approach in the limit $E\rightarrow 0$.
$A$ and $B$ are both triply degenerate while $C$ is doubly degenerate because
of its time reversed partner.
However, in the fundamental domain all three orbits are non-degenerate.
Orbit $A$ fluctuates between being stable and unstable
as the energy is varied.
Orbit $B$ is unstable for all energies and orbit $C$ is stable for all
energies up to $E\approx 0.145$ where it suffers a tangent bifurcation.
In Fig.~3, we show the periods $T$ as well as the stability angle $v$
or exponent $u$ for the two orbits $B$ and $C$.
The stability angle of the unstable orbit $B$ is a constant $v=4\pi$,
corresponding to a Maslov index 4.

The ``rotational greens function'' $g_\phi(E)$ probes the class
of rotations and of the three short orbits only has contributions from $C$.
Each third multiple modulus 1 or 2 of repetitions of the primitive orbit
in the fundamental domain contributes to the density, and we obtain
\begin{equation} \label{dg(E)}
g_\phi(E) =
\frac{T}{i\hbar}\sum_{k=0}^{\infty}\sum_{r=0}^{\infty}\left(
e^{i\alpha({1\over 3}+r)} + e^{i\alpha({2\over 3}+r)} \right)
\end{equation}
where $\alpha=S/\hbar-v(k+1/2)$.
Since long orbits will be damped by the smoothing parameter $w$,
we limit ourself to $r=0$.
After summing over $k$ we obtain the result
\begin{equation} \label{dgsimp}
g_\phi(E) \approx -\frac{T}{2\hbar}\left(\frac{e^{iS/3\hbar}}{\sin v/6}
+ \frac{e^{2iS/3\hbar}}{\sin v/3}\right).
\end{equation}

We next convolve this expression with the Gaussian in order to compare with
the exact spectrum which has been similarly convolved.
To leading order in $\hbar$ we obtain for the smoothed ``rotational
density''
\be \label{finex}
\tilde{\rho}_{\phi ,{\rm osc}}(E) \approx \frac{T}{2\pi\hbar}
\left(\frac{\sin(S/3\hbar)}{\sin v/6}e^{-T^2w^2/18\hbar^2} +
\frac{\sin(2S/3\hbar)}{\sin v/3}e^{-4T^2w^2/18\hbar^2}\right) .
\ee
The correction to this formula is of order
$\mbox{O}(e^{-16T^2w^2/18\hbar^2})$
and comes from higher repetitions of the orbit.
Clearly, with the use of a Gaussian damping we have a strong suppression
of the longer orbits and higher repetitions of short orbits.

The oscillatory terms may diverge for critical values of the stability
angle $v$. The energy dependence of $v$ is shown in Fig.~3.
At zero energy it has the value of $4\pi$ and approaches $3\pi$ near a
critical energy close to $0.145$ at which
point the orbit becomes unstable and $v$ becomes complex.  The denominator of
the second term of Eq.~(\ref{finex}) is zero when $v=3\pi$ so the
corresponding contribution diverges.
In this case, we should go to the next
order in the saddle point approximation as discussed in Ref.~\cite{kus}.
However, since the stability angle approaches zero abruptly near the
critical energy, this divergence is only significant for energies very close
to the critical energy.

The final step is to integrate Eq.~(\ref{finex}) with respect to energy to
get the spectral staircase function. In principle,
$T$, $S$ and $v$ all depend on energy but to a good approximation we can
assume $S\approx TE$ and ignore the energy dependence of $T$ and $v$ in the
integral to obtain the final expression
\begin{equation} \label{lastex}
\tilde{N}_{\phi,{\rm osc}} (E)\approx -\frac{3}{2\pi}
\left(\frac{\cos{(S/3\hbar)}}{\sin{v/6}}e^{-T^2w^2/18\hbar^2} +
\frac{\cos{(2S/3\hbar)}}{\sin{v/3}}e^{-4T^2w^2/18\hbar^2}\right).
\end{equation}

In Fig.~4 we show the results of the exact smoothed density of states and the
results of formulas (\ref{res2}) and (\ref{lastex}) for two values of the
smoothing.
For $w=0.004$, it is clear that the first term of Eq.~(\ref{lastex})
is sufficient to describe the oscillations although both terms were included.
For $w=0.0015$, inclusion of the second term is essential.
In both cases there is very good agreement over the entire energy scale
shown.
The agreement in Fig.~4 is remarkable given the usual
inadequacies of the trace formula in mixed systems.
We will see that for $N_\sigma(E)$ the agreement is much poorer.
It is worth mentioning that the good
agreement for $N_\phi (E)$ is in contradiction to the result of
Ref.~\cite{badback}.
There the authors found that, for unknown reasons, the theory
describes their numerical results better if the multiplicity of orbit $C$ is
taken
to be three rather than two.  Had we done the same, we would have found a
fifty percent discrepancy in the amplitude of the oscillations rather than the
near perfect agreement shown.

The oscillations in the other combination, $N_\sigma (E)$, are dominated by the
unstable orbit $B$.  A similar analysis to that for orbit $C$ gives the result
\begin{equation} \label{ultex}
\tilde{N}_{\sigma,{\rm osc}}(E) \approx -\frac{\sin{S/2\hbar}}
{\pi\sinh{u/4}}e^{-T^2w^2/8\hbar^2},
\end{equation}
where $u$ is the stability of the orbit. The action $S$ and the period
$T$ are close, but not identical, to those of orbit $C$.
Once again, these are defined as the values
on the full configuration space.  The function $u(E)$ is plotted in the bottom
of Fig.~3. Eq.~(\ref{ultex}) only contains the
contribution from a single repetition. The correction comes from the boundary
orbit $A$ and higher repetitions of $B$
and is of order $\mbox{O}(e^{-T^2w^2/2\hbar^2})$.  The result is shown in
Fig.~5 where, unlike before, the agreement is quite poor except in a small
region of energy between about $0.10$ and $0.16$. This can be explained by the
fact that the expression diverges when $u$ is zero.  However, as is apparent
from Fig.~3, $u$ is small for a large range of energy.  This is in contrast to
the behaviour of $v$ which is close to its singular value of $3\pi$ only in a
narrow range of energy. Furthermore, the divergence in Eq.~(\ref{lastex})
is in the suppressed second term and not in the leading term as in
Eq.~(\ref{ultex}).

\section{Conclusion}

In this paper, we have developed a consistent scheme for the evaluation of
the Weyl density of states for potential problems in the presence of either
discrete or continuous point group symmetries.
For discrete groups, our formalism allows us to express the density of states
of any irrep as a specific combination of contributions from the various
classes.
By decomposing the orthogonal transformations into a product of
reflections and rotations, we were able to evaluate the contribution coming
from any conceivable class of group elements.
Each class contributes with a relative power of $\hbar$ which
is simply the codimension of the set of points which is invariant with
respect to that group element.
For an $n$ dimensional system, we are usually interested in evaluating at least
the first $n+1$ terms of the Weyl series so that the remainder decreases with
$\hbar$. Because of this, it is important to have a scheme to find
the $\hbar$ expansion of the contribution from a specific class and also
to be able to find the contributions from all possible classes. In
this paper, we have shown how both of these can be done.

As an example, we showed how these expansions can be used to evaluate the
Weyl density of states for the two dimensional H\'{e}non-Heiles potential and
excellent
agreement was found. In addition, we evaluated the contributions from the first
periodic orbits and also found good agreement with the numerics - again within
the symmetry reduced picture.

The calculation involving continuous rotational groups is subtle because of
the fact that the limits $\hbar\rightarrow 0$ and $\theta\rightarrow 0$
do not commute. For $SO(2)$ states, we were able to evaluate the contribution
of all rotations to the $m=0$ states and found that these states can be
identified exactly with the $L=0$ manifold in the classical phase space. For
other $m$ states, the identification is not so clean and we can not evaluate
$\rho_m(E)$ exactly but rather must do a local average over the index $m$.
We get a similar result for $SO(3)$ and a simple extension will
give the result for any group of the form $SO(n)$.

In this work, the discussion was for Hamiltonians of the form $H=T+V$ where
$T$ depends only on the momenta and $V$ depends only on the coordinates.
This encompasses the majority of physical systems in which one is interested.
However, one is sometimes interested in more general situations where terms
involving momenta and coordinates are mixed.  One example is a particle in a
magnetic field, which has the quantum Hamiltonian
\be
\hat{H} = {1 \over 2}\left(\hat{\bf p}-
e{\bf A}(\hat{{\bf q}})\right)^2
+ V(\hat{{\bf q}}).
\ee
Another example is the class of systems for which the dynamics is defined in
terms of group
generators, such as the kicked top \cite{kt} or various algebraic models
\cite{almod}. In these
systems, the designation of what is momentum and what is coordinate is
arbitrary and the Hamiltonian is composed of mixed terms. It is possible to
extend the results of this paper to such systems. However, this is more
complicated than what we have done since
the Wigner transform of the
Hamiltonian is no longer found by simply replacing the quantum momentum and
coordinate operators by their classical counterparts as in
Eq.~(\ref{wighamm}). Doing so yields only the first term in an
$\hbar$-expansion of the Wigner transform of the Hamiltonian operator.
In principle, we would need to
evaluate all of the terms and keep track of this expansion throughout the
remainder of the calculation.

We would like to thank the authors of Ref.~\cite{badback} for the use
of Fig.~2. In addition, we would like to thank Rajat Badhuri for interesting
discussions. This work was supported by the EU Human Capital and Mobility
Fund and by the Carlsberg Foundation.

\newpage
\begin{table} \label{tab1}
\begin{center}
\begin{tabular}{rrrr}
  & $I$ & $C_3^{(2)}$ & $\sigma_v^{(3)}$\\
\hline
$A_1$ \vline & $1$ &  $1$         &   $1$         \\
$A_2$ \vline & $1$ &  $1$         &  $-1$         \\
$E$ \vline & $2$ & $-1$         &   $0$
\end{tabular}
\caption{Character table of the group $C_{3v}$.  The superscripts on the
class labels indicate the number of group elements in that class.}
\end{center}
\end{table}

\begin{figure}
\caption{The top box shows $N_\phi (E)$ and the bottom box shows $N_\sigma (E)$
as defined in the text. Both refer to the the H\'{e}non-Heiles
potential.  The solid lines are the exact numerics while the dashed lines
are the analytical estimates.
Both of these are for $\hbar=.004$ with a smoothing width $w=0.01$.}
\end{figure}

\begin{figure}
\caption{The configuration space. The lower left triangle depicts the
fundamental domain.
The solid lines show the three fundamental periodic orbits as
they exist in the fundamental domain. The dashed lines show the same orbits
in the full configuration space.}
\end{figure}

\begin{figure}
\caption{The top box shows the period $T(E)$ of the two periodic
orbits $B$ (dashed line) and $C$ (solid line).
The middle section gives
the stability angle $v(E)$ for periodic orbit $C$.
It varies between $4\pi$ at $E=0$ and $3\pi$ at $E\approx 0.145$,
as indicated by the dotted lines. The bottom
box shows the stability exponent of periodic orbit $B$.}
\end{figure}

\begin{figure}
\caption{$N_\phi (E)$ for two smaller values of the smoothing width,
$w=.004$ in the
upper figure and $w=.0015$ in the lower figure. The solid lines are the exact
numerics while the dashed lines are the analytical estimates including two
repetitions of periodic orbit $C$.}
\end{figure}

\begin{figure}
\caption{$N_\sigma (E)$ for a smaller value of the smoothing, $w=.0015$. The
lower figure is simply an expansion of the region between $E=0.10$ and
$E=0.16$.   The solid lines are the exact numerics while the dashed lines
are the analytical estimates including one repetition of periodic orbit
$B$.}
\end{figure}

\end{document}